\documentclass[twocolumn,aps,footinbib,showpacs,showkeys,preprinttightenlines,superscriptaddress,preprintnumbers] {revtex4}

\usepackage{amssymb}
\usepackage{amsfonts}
\usepackage{amsmath}
\usepackage{bm}
\usepackage{color}
\usepackage{slashed}
\usepackage{graphicx}
\usepackage{color}
\usepackage{epsfig}
\usepackage{hhline}
\usepackage{longtable}
\usepackage{multirow}
\usepackage{rotating}
\usepackage{hyperref}
\usepackage{color}

\newcommand{\dis}[1]{\begin{equation}\begin{split}#1\end{split}\end{equation}}
\newcommand{\be}{\begin{equation}}
\newcommand{\ee}{\end{equation}}
\def\bea{\begin{eqnarray}}
\def\eea{\end{eqnarray}}

\newcommand{\eq}[1]{Eq.~(\ref{#1})}

\newcommand{\bfrac}[2]{{\left(\frac{#1}{#2} \right)  }}\newcommand{\VEV}[1]{\langle #1 \rangle}

\newcommand{\Mp}{M_P}

\newcommand\gev{\,{\rm GeV}}
\newcommand\mev{\,{\rm MeV}}
\newcommand\kev{\,{\rm keV}}
\newcommand\ev{\,{\rm eV}}

\newcommand\mpc{\,{\rm Mpc}}

\newcommand\mw{m_{\rm W}}
\newcommand\ms{m_{\rm s}}

\newcommand\treh{{\rm T}_{\rm R}}

\newcommand{\Neff}{N_{\rm eff} }
\newcommand{\NeffSM}{N_{\rm eff}^{\rm SM} }
\newcommand{\NeffBBN}{N_{\rm eff}^{\rm BBN} }
\newcommand{\NeffCMB}{N_{\rm eff}^{\rm CMB} }

\newcommand{\rhoDR}{\rho_{\rm DR} }

\newcommand{\lamf}{\lambda_{\rm f} }

\def\bea{\begin{eqnarray}}
\def\eea{\end{eqnarray}}
\definecolor{brown}{rgb}{0.5,0.2,0.0}

\begin{document}
\title{Dark radiation and small-scale structure problems with decaying particles}

\author{Kiwoon Choi}
\email{kchoi@kaist.ac.kr}
\affiliation{Physics Department, Korea Advanced Institute of 
Science and Technology Daejeon 305-701, Republic of Korea}

\author{Ki-Young Choi}
\email{kiyoung.choi@apctp.org}
 \affiliation{Asia Pacific Center for Theoretical Physics, Pohang, Gyeongbuk 790-784, Republic of Korea}
  \affiliation{Department of Physics, POSTECH, Pohang, Gyeongbuk 790-784, Republic of Korea
}

\author{Chang Sub Shin}
\email{csshin@apctp.org}
\affiliation{Asia Pacific Center for Theoretical Physics, Pohang, Gyeongbuk 790-784, Republic of Korea}

\color{black}
\begin{abstract}
Although the standard $\Lambda$CDM model describes the cosmic microwave background radiation and the large scale structure 
of the Universe with great success, it has some tensions with observations in the effective number of neutrino species (dark radiation) and the number of small scale structures (overabundance problem).  
Here we propose a scenario which can relax these tensions by producing both dark matter and dark radiation by late decays of heavy particle. Thanks to the generation mechanism, dark matters are rather warm so that the
small-scale structure problem is resolved. This scenario can be naturally realized in supersymmetric axion model, in which axions produced by saxion decays provide dark radiation, while axinos from saxion decays form warm dark matter. We identify a parameter region of supersymmetric axion model satisfying all known cosmological constraints.

\end{abstract}


\preprint{APCTP Pre2012 - 012}

\maketitle

\section{Introduction}

The standard $\Lambda$CDM cosmological model has been extremely successful in explaining the  observed acoustic peak in the cosmic microwave background (CMB) radiation and the formation of large scale structures (LSS). Despite its success, the $\Lambda$CDM model seems to have some tensions
with observations at small scales.
The measurement of the temperature anisotropy of the CMB showed less power spectrum at small scales, suggesting that the number of effective neutrino species, $\Neff$, has 
a bigger value  than the one  
predicted by the standard model of particle physics, so the existence of `dark radiation'.
Another difficulty of $\Lambda$CDM model is faced at small scales of the structure formation. N-body simulation 
with cold dark matter (CDM) has shown a tension with observation
in the nonlinear regime of  structure formation, producing  more substructures in  Milky-Way galaxy size than the observed ones.

In the standard cosmological scenario, the thermal plasma after the electron-positron annihilation
contains photons and neutrinos. 
At this epoch, 
total radiation energy density  can be parameterized as 
\dis{
\rho_{\rm rad}= \left[ 1+ \Neff \frac78 \bfrac{T_\nu}{T_\gamma}^4 \right]\rho_\gamma,
\label{Neff}
}
where $\rho_\gamma=(\pi^2/15)T_\gamma^4$ is the photon energy density, 
$T_\nu/T_\gamma=(4/11)^{1/3}\simeq 1.40$ after the electron-positron annihilation, 
and $\Neff$ is the effective number of neutrinos, including the contribution
from dark radiation if there exists any. In the standard model with three neutrino flavors, the residual heating of the neutrino fluid due to the electron-positron annihilation slightly increases $\Neff$, yielding   $\Neff^{\rm SM}= 3.046$~\cite{Mangano:2005cc}. 
However, the WMAP collaboration reported  $\Neff=4.34^{+0.86}_{-0.88} \,(68\% {\rm CL})$ through  the measurements of Hubble constant and baryon acoustic oscillation~\cite{Komatsu:2010fb}. Similarly higher values of $\Neff$ are observed by Atacama Cosmology Telescope (ACT) and South Pole Telescope (SPT), reporting
$\Neff = 4.56\pm 0.75$~\cite{Dunkley:2010ge} and $\Neff= 3.86\pm 0.42$~\cite{Keisler:2011aw}, respectively. 
It is expected that the Planck satellite will be able to measure
$\Neff$ with better precision~\cite{Hamann:2007sb}, so make the 
situation more clear.

The $\Neff$ measured in the CMB, $\NeffCMB$, can be compared with the value $\NeffBBN$ determined by the big bang nucleosynthesis (BBN). Observations of the primordial ${}^4{\rm He}$ abundance provides the best constraint on $\NeffBBN$. However there is a controversy between different groups about the relic helium abundance, e.g.
\cite{Izotov:2011pa} and 
\cite{Aver:2010wq}, while another recent analysis by Mangano and Serpico~\cite{Mangano:2011ar} gives an upper bound $\NeffBBN \leq 4 \, (95\% {\rm CL})$.
At any rate, a larger value of $\Neff$ can be explained by
extra relativistic degree of freedom existing at the epoch prior to the recombination. Many models are suggested to explain this dark radiation~\cite{Neff models}, 
including the ones considering  the decays of heavy particle as the origin of dark radiation~\cite{Neff decays}.  

In the N-body simulation with CDM, the structures form hierarchically, with small structures collapsing first and merging into larger and larger bodies. CDM model describes the distribution and correlation of structures very well at large scales, however there is a large discrepancy at small scales
between the observed number of satellite galaxies of the Milky Way and the expected number~\cite{Klypin:1999uc}. This tension has brought many questions on the galaxy formation and evolution as well as the properties of dark matter.
One possibility is to impose warm nature of dark matter (WDM) instead of coldness~\cite{Bode:2000gq}.  
The free-streaming of WDM can reduce the power spectrum at small scales,
which would  result in smaller number
of galactic 
subhalos~\cite{WDM}.

In fact, WDM model with $\mw\simeq 1 - 4\kev$ can alleviate the CDM overabundance problems in many respects. It resolves the discrepancy in the bright satellite galaxies~\cite{Lovell:2011rd}, solves the excess of predicted faint galaxies at low and high redshifts, as well as the excess of bright galaxies at low redshifts in the galaxy formation~\cite{Menci:2012kk}. Also it has 
better agreement in the HI velocity (width) function measured in the ALFALFA survey~\cite{Papastergis:2011xe}, and 
in the number of Milky Way satellites~\cite{Polisensky:2010rw}.

In this work, we examine a scenario
in which  both dark radiation and WDM
find a common origin in
the decays of  heavy particle.
As we will see, supersymmetric axion model with relatively 
light saxion and axino masses provides a natural set up
realizing such scenario. 
Decays  of massive saxion produce  
(nearly) massless axion pairs and massive axino pairs
with different branching ratios. Axions then 
contribute to the dark radiation, while
axinos  become warm dark matter which have large
velocities to solve the small scale structure problems.

\section{Dark radiation and warm dark matter from particle decays}

Let us consider a non relativistic particle $X$, which decays dominantly to a pair of light particles (${\rm DR}\equiv$ dark radiation)
 and also to a pair of massive particles  of 
 mass $m$ (${\rm DM}\equiv$ dark matter)
 with small branching ratio.
The energy density of nonrelativistic particles decreases as $a^{-3}$, while the radiation energy density behaves as $a^{-4}$. Therefore
even when the mass density of $X$ was initially subdominant, it can be important when $X$  decays.
After decay,  all daughter particles are relativistic, however massive ones become nonrelativistic later 
due to the redshift of the momentum.
After $X$ decays, the resulting dark radiation energy density 
can be parameterized by 
the extra effective number of neutrino species  
$\Delta \Neff\equiv \Neff-\NeffSM$, which is given by
\dis{
 \Delta \Neff (t)=\NeffSM \frac{\rho_{\rm DR}(t)}{\rho_\nu(t)}=   \bfrac{8}{7} \bfrac{11}{4}^{4/3}  \frac{\rho_{\rm DR}(t)}{\rho_\gamma(t)},
}
where $\rho_\nu=\frac78 \NeffSM T_\nu^4$ and $\rhoDR$ is the extra relativistic energy density called dark radiation.  

Here and in the followings, we use the instantaneous decay 
approximation, and  assume that the branching ratio 
of the $X$ decay into DM pair is small enough, so the DM density at the 
time $\tau_X$ of $X$ decays is negligible.
Then the DR energy density right after $\tau_X$ 
is
nearly equal to $\rho_X$ right before $\tau_X$:
\dis{
\rho_{\rm DR} (\tau_X) = \rho_X(\tau_X) = s(\tau_X) M_X Y_X ,
}
where  $M_X$ and $Y_X\equiv n_X/s$ are the mass and the abundance  of the decaying $X$. Using the entropy density $s =2\pi^2/45 g_{*S}T^3$  with $g_{*S}\simeq 3.91$ for $T\lesssim 1 \mev$, one easily finds that
$\Delta \Neff$ at time $\tau_X$  is given by
\dis{
 \Delta \Neff (\tau_X) &= \bfrac{8}{7} \bfrac{11}{4}^{4/3}  \frac{ s(\tau_X)}{\rho_\gamma(\tau_X)}M_X Y_X\\
 &\simeq 11.5 \bfrac{1\kev}{T_\gamma(\tau_X)}\bfrac{M_X Y_X}{1\kev}, \label{DNeffT}
}
where the relation between lifetime and the temperature in the radiation-dominated epoch is given by
\dis{
\tau_X\simeq \bfrac{90}{\pi^2 g_*}^{1/2} \frac{\Mp}{T_\gamma^2}\simeq 2.6\times 10^6  \bfrac{1\kev}{T_\gamma}^2 \sec,
}
with the reduced Planck mass $\Mp=2.4\times 10^{18}\gev$  and $g_* \simeq 3.36$.
Then \eq{DNeffT} can be reexpressed as
\dis{
 \Delta \Neff (\tau_X) \simeq 7.1 \bfrac{\tau_X}{10^6\sec}^{1/2}\bfrac{M_X Y_X}{1\kev}.
 \label{DNeff}
}

\begin{figure}[!t]
 \begin{center}
  \begin{tabular}{c}
   \includegraphics[width=0.45\textwidth]{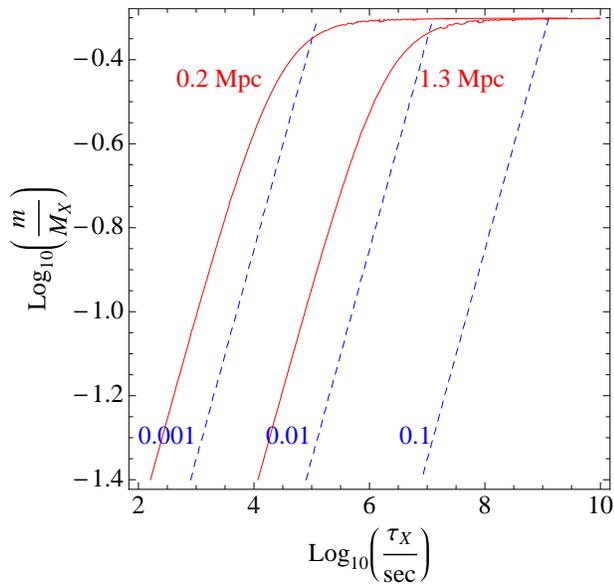}
  \end{tabular}
  \end{center}
  \caption{Blue dashed lines  show the contour plot of $c_mf_m=0.1,0.01,0.001$  which gives $\Delta \Neff (\tau_X)=1$ when dark matters are produced from the decay of $X$ (equality in \eq{condition1}).  Red solid lines are the contour plot of $\lambda_{\rm FS}=0.2, 1.3 \, \mpc$ in the plane of $m/M_X$ and $\tau_X$.
  In the region between the red lines and  $f_m \sim 0.001$,  both the dark radiation and small scale structure problems can be explained with the decay of particles.}
  \label{fig:Neff1}
\end{figure}

The DM particles produced from the decay of $X$  are relativistic initially. However their momenta are red-shifted due to the expansion of the Universe, making them 
non-relativistic at the epoch
\dis{
t_{\rm NR} \simeq \bfrac{M_X}{2m}^2 \tau_X,\label{tNR}
}
when the red-shifted momentum becomes comparable to the mass 
\dis{
p(t_{\rm NR}) = \frac{M_X}{2}\bfrac{\tau_X}{t_{\rm NR}}^{1/2} \simeq m. 
}
After this epoch, the energy density of the DM particles 
produced by the decays of $X$ decreases more slowly than the radiation
energy density, and constitute  the non-thermally produced
dark matter mass density: 
\dis{
\Omega^{\rm NTP}_{\rm DM} h^2 &=2f_{m} \frac{m}{M_X}\Omega_X h^2\\
&= 5.4 \times 10^{10}\,f_m \bfrac{m}{100\gev}Y_X, \label{NTP}
}
where $f_m$ is the branching fraction of
$X\rightarrow {\rm DM} +{\rm  DM}$.
Imposing the condition 
\dis{
\Omega^{\rm NTP}_{\rm DM} h^2 \leq \Omega^{\rm WMAP}_{\rm DM} h^2 =0.11, \label{WMAP}
}
the mass ratio between  $m$ and $M_X$ is constrained as
\dis{
f_m\frac{m}{M_X} \leq 2\times 10^{-4} \bfrac{1\kev}{M_XY_X},
}
or equivalently (using \eq{DNeff})
\dis{
f_m\frac{m}{M_X} \leq  1.4 \times 10^{-3} \frac{1}{\Delta \Neff (\tau_X)} \bfrac{\tau_X}{10^6 \sec}^{1/2}. 
\label{condition1}
}
In figure~\ref{fig:Neff1}, we show the contour plot of $f_m=0.1,0.01,0.001$ (blue sold lines) in the plane of $m/M_X$ and $\tau_X$
which gives $\Delta \Neff (\tau_X)=1$ assuming that most of the dark matters are produced from the decay of $X$.

Then, using \eq{tNR} we find the time when DM becomes non relativistic is
given by 
\dis{
t_{\rm NR} \geq 12.8 \times 10^{10}  \, ( f_m  \Delta  \Neff(\tau_X) )^2 \sec,
} 
and therefore the DM particles produced by the decays of $X$  
can have large kinetic energy which can erase the small scale structure formation.
The characteristic free-streaming length is given by
\dis{
\lambda_{\rm FS} = \int^{t_{\rm eq}}_{\tau_X} \frac{v(t)}{a(t)} dt,
}
which can be approximated as
\dis{
\lambda_{\rm FS} \simeq 1.0 \, {\rm Mpc} \, \bfrac{u_\tau^2 \tau_X}{10^6 \sec}^{1/2} \left[ 1-0.07 \ln\bfrac{u_\tau^2 \tau_X}{10^6 \sec} \right],
}
where $u_\tau$ is evaluated at  $\tau_X$ and expressed as
\dis{
u_\tau \equiv \frac{|\vec{p}|}{m}\simeq&\frac{M_X}{2m}\left(1- 4\frac{m^2}{M_X^2} \right).
}
The characteristic free-streaming length can be related to the thermally produced warm dark matter mass via~\cite{SommerLarsen:1999jx}
\dis{
\lamf \equiv \frac{2\pi}{k_f}= 1.29 \bfrac{\Omega_m h^2}{0.11}^{1/3} \bfrac{m_{\rm W}}{\kev}^{-4/3}\, \mpc.
\label{lambdaf}
}

The Lyman-$\alpha$ forest data constrains the cut-off scale of the power spectrum. In terms of the warm dark matter mass $m_{\rm W}$, it has been claimed to give a $2\sigma$-bound $\mw> 2\kev $~\cite{Viel:2006kd,Seljak:2006qw}, however it can be relaxed to $\mw>0.9\kev$ if the less reliable data are rejected~\cite{Viel:2006kd,Boyarsky:2008xj,Aarssen:2012fx}. 
Considering the blazer heating,
the revised  bound $\mw > 1.7\kev$~\cite{Boyarsky:2008ju} 
can have about $30 \%$ systematic uncertainty~\cite{Puchwein:2011sx}. 
Therefore here we adopt WDM with $\mw= 1 - 4 \kev$ as a solution for the small scale structure formation, which can be consistent with
the Lyman-$\alpha$ constraint.
This corresponds to the free streaming scale $\lambda_{\rm FS}= 0.2 - 1.3 \mpc$.
In Fig.~\ref{fig:Neff1}, we show the contour plot of $\lambda_{\rm FS}=0.2, 1.3 \, \mpc$ in the plane of $\tau_X$
and  $m/M_X$.

\section{Supersymmetric axion model}
Supersymmetric axion model provides a viable example 
which  would realize the scenario 
 discussed in the previous section. 
The model includes a $U(1)_{\rm PQ}$ symmetry spontaneously broken by the vacuum expectation value of a PQ charged but SM neutral scalar field $\phi$. We can identify the radial component of $\phi$ as the saxion  
$s$, and the phase component as  the axion $a$. If there
exists SU(3)$_C$ charged fermion which transforms under $U(1)_{\rm PQ}$,  
the associated $a$ becomes the QCD axion 
solving  the strong CP problem~\cite{Kim:2008hd}. 
In supersymmetric model, $\phi$ can be considered as a chiral superfield
given by
\bea
S=  \left(F_a + \frac{s}{\sqrt{2}}\right)\exp\left(\frac{i a}{\sqrt{2}F_a}\right) 
+ \sqrt{2}\theta\tilde a + \theta^2 F^S,
\eea
where $F_a=\VEV{\phi}$ is the axion decay constant, 
and $\tilde a$ is the axino, the fermionic superpartner of axion. 
While the axion gets a mass only by QCD anomaly, the saxion and axino can be much 
heavier than the axion due to supersymmetry (SUSY) breaking effects.  Because the interactions
of the axion supermultiplet  are suppressed by the axion decay constant, 
one can easily find a setup which would explain the dark radiation and small scale problems  with saxion decaying mostly to
axions for dark radiation and also to axinos with  small branching ratio 
for warm dark matter.

The relevant Lagrangian for our discussion is as follows.
\dis{
\label{Axion_interaction}
{\cal L}&=\frac{1}{2}(\partial_\mu s)^2 + \frac{1}{2}(\partial_\mu a)^2 +
\frac{1}{2}\bar {\tilde a}\ i \bar \sigma^\mu\partial_\mu \tilde a \\ &
-\frac{1}{2}\ms^2s^2   -\frac{1}{2}\Big( m\tilde a \tilde a + h.c.\Big)
\\ &+\frac{s}{\sqrt{2}F_a}\left[
(\partial_\mu a)^2 +\frac{\lambda}{2} \Big(m\tilde a \tilde a + h.c.\Big) 
+\sum_{A}\frac{ g_A^2C_A}{32\pi^2} F^{A\mu\nu} F^A_{\mu\nu}\right].
}
Here $\ms$ and $m$ denote the saxion and axino masses, respectively,
and $\lambda$ and $C_A$ are model-dependent parameters of order unity or smaller than 
one. The axion mass is neglected in our discussion, because we assume that  
the axion has a very small mass  as that of the usual QCD axion, 
$m_a\sim 6\times 10^{-6}\ev (10^{12}\gev/F_a)$.
For the KSVZ type axion model~\cite{Kim:1979if}, the above terms are enough, while 
 for the DFSZ type axion model~\cite{Dine:1981rt}, saxion can have
 sizable couplings to the SM fermions which are charged under 
 $U(1)_{\rm PQ}$.
 Here we take the KSVZ type model as our example, in which 
 the SM fermions are not PQ charged, and so their couplings to saxion and
 axion can be ignored.

In the limit $\ms \gg m$, saxions decay dominantly to axion pairs with 
a decay rate
\dis{
\Gamma(s\rightarrow 2a) \simeq \frac{1}{64\pi}\frac{\ms^3}{F_a^2},
}
yielding the saxion lifetime
\dis{
\tau_s \simeq 1.3\times 10^5  \bfrac{\ms}{100\mev}^{-3}\bfrac{F_a}{10^{12}\gev}^2 \sec.
}
The saxion field is initially displaced from the present vacuum value, and starts oscillation at the moment when
the expansion rate $H\sim \ms$. If it happens before the reheating after inflation, the energy density to entropy density ratio, which is constant during the radiation-dominated epoch,
is given by
\dis{
&\frac{\rho_s}{s} = \ms Y_s = 2.2 \times 10^{-8}  \bfrac{\treh}{10^6\gev} \bfrac{F_a}{10^{12}\gev}^2 \, \gev,
}
where we used the initial displacement of the saxion $\delta s\simeq F_a$.
We then find from \eq{DNeff} that
$\Delta \Neff$ at the time of saxion decay is given by
\dis{
\Delta \Neff=0.056 \left(\frac{100\mev}{\ms}\right)^{3/2}
\left(\frac{F_a}{10^{12}\gev}\right)^3 \left(\frac{\treh}{10^6\gev}\right). \label{Neffaxion}
}

\begin{figure}[!t]
  \begin{center}
  \begin{tabular}{c}
   \includegraphics[width=0.45\textwidth]{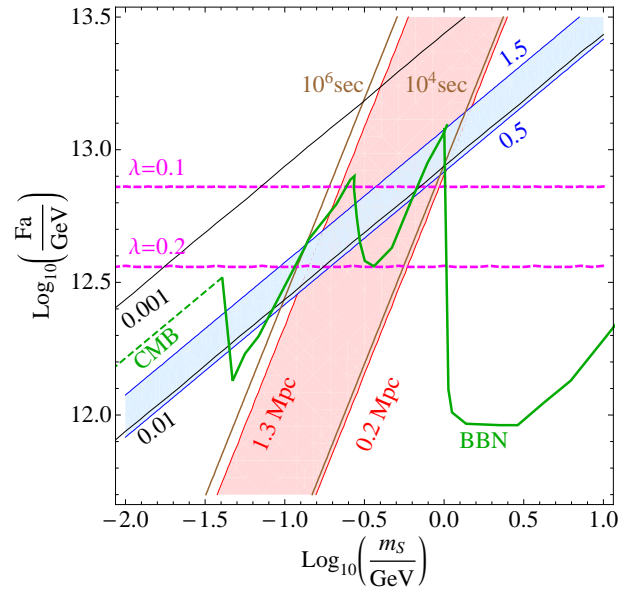}
  \end{tabular}
  \end{center}
  \caption{Contour plots of $\Delta \Neff$ and $\lambda _{\rm FS}$ in the ($\ms$,$F_a$) plane with other cosmological constraints.  Here we used $T_R=5\times 10^5\ {\rm GeV}$, $m/\ms=0.25$.
  Blue lines denote $\Delta N_{\rm eff}=0.5,\, 1.5$.
  Red lines show $\lambda_{\rm FS}=0.2, \, 1.3\mpc$.
 Black lines correspond to thermal production of the axino 
 $\Omega_{\tilde a}^{\rm TP}h^2=0.001,\, 0.01$. 
 Brown lines denote the lifetime of the saxion $10^4\sec,\, 10^6\sec$.
  The horizontal magenta (dotted)  lines represent
 $\Omega_{\tilde a}^{\rm NTP}h^2=0.1$ for $\lambda=0.1, 0.2$ respectively. 
 The green line shows the BBN (solid) and CMB (dashed) constraint  and  the lower region is allowed.}
  \label{fig:Neff2}
\end{figure}

The saxion also produces axinos with decay rate
\dis{
\Gamma(s\rightarrow \tilde a\tilde a) = \frac{\lambda^2 m^2 \ms }{32\pi F_a^2} \left[ 1- \bfrac{4m^2}{\ms^2}\right]^{3/2},
}
for which the branching ratio is given by 
\dis{\label{branching_ratio}
f_m\simeq \left(\frac{2\lambda^2m^2}{\ms^2}\right) \left(1-\frac{4 m^2}{\ms^2}\right)^{3/2}.}
Such non-thermally produced axinos play the role of warm dark matter and can solve the small scale structure problems as explained in the previous section.

In Fig.~\ref{fig:Neff2}, we show the viable region in the plane of $\ms$  and $F_a$  for  $m/\ms =0.25$ and the
reheat temperature  $\treh=5\times10^5\gev$ of the primordial inflation. 
The blue lines denote $\Delta N_{\rm eff}=0.5$ and 1.5,
while the red lines stand for $\lambda_{\rm FS}= 0.2$ and  $1.3 \mpc$. On the dashed Magenta line, axinos produced by saxion decays
constitute most of the dark matter for $\lambda =0.1$ and  0.2. 
Therefore in the overlapped region of blue and red bands, 
which corresponds to $100\mev \lesssim \ms \lesssim 1\gev$ and $3\times10^{12}\gev \lesssim F_a \lesssim 10^{13}\gev$, both dark radiation and small scale structure problems can be explained with corresponding value of $\lambda$ between $0.1$ and $ 0.2$ respectively.
In this region the relic density of the thermally produced  axinos is less than about  10 \% of dark matter (black solid line), thus most of the axino dark matters are produced from saxion decays. 

The decay of saxions can produce electromagnetic and hadronic particles which can disrupt the light element abundances after BBN.  The lifetime of saxion in the region of our interest is between $10^4$ and $10^6 \sec$.
 In Fig.~\ref{fig:Neff2},   we show the BBN and CMB bound with green lines~\cite{Jedamzik:2004er} and the upper region is disallowed.
In this region, the constraints on the hadronic and electromagnetic injections lead to
\dis{
B_{\rm h} \frac{\rho_s}{s} \lesssim 10^{-14} \gev,
\quad
B_{\rm em} \frac{\rho_s}{s} \lesssim 10^{-6} - 10^{-13} \gev,
} 
where $B_{\rm h}$ and $B_{\rm em}$ denote the branching ratios
for the hadronic and electromagnetic injections.
These constraints can be easily satisfied if the saxion mass is below
the pion production threshold $\ms < 2m_\pi $.

For the axino mass much smaller than the value of  $ {\cal O}(0.1) \ms$,
the branching fraction (\ref{branching_ratio}) might be too small to provide
the correct amount of DR and WDM simultaneously, for a given $\lambda$,
as presented in Fig.~\ref{fig:Neff1}. 
 If we take a rather large coupling, $\lambda \gg 1$,  
the right amount for DR and WDM can be obtained for  high reheating temperature $\treh$. 
However, in this case, 
the thermal production of  axino becomes much larger than the non-thermal production 
 from saxion decays~\cite{foot}.
Consequently, the mass range $m \ll {\cal O}(0.1) \ms$ is not favored by our scenario. 
The  cosmological and astrophysical constraints on supersymmetric axion models  are also well summarized  in \cite{Kawasaki:2007mk}.

We then find that the saxion mass around $100\mev $ with Peccei-Quinn scale around $5\times 10^{12}\gev$
and reheating temperature $5\times 10^5\gev$ can explain both dark radiation and small scale structure problems, while satisfying all the constraints from BBN, cold axino abundance, and gravitino problem.
If the axion in our model is a QCD axion solving the strong CP problem, 
we need a small axion misalignment angle $\theta_i \lesssim 0.1$ 
in order for  the cold axion dark matter
to be subdominant compared to the warm axino dark matter produced by saxion
decays.


We close this section with a brief discussion of SUSY breaking schemes
which can give the saxion mass $\ms\sim 100\mev$ and the axino mass
$m\sim 0.2 \ms$.
We first note that in gauge mediation
with a messenger scale $M_{\rm mess}\ll F_a$, saxion and axino masses
can be much lighter than the MSSM soft parameters as they 
are further suppressed by some powers
of $M_{\rm mess}/F_a$.
Furthermore if saxion is stabilized by radiative effects, a small hierarchy
between the saxion and axino masses, e.g. $m/\ms\sim \sqrt{1/4\pi^2}$,
can arise in a natural manner.
On the other hand, if there exists a SUSY breaking sector well sequestered
from the visible sector as well as from the PQ sector,
the gravitino mass can be much heavier than the axino mass.
These points suggest that there can be a plenty of rooms for the mediation
of SUSY breaking yielding the desired saxion and axino masses while
satisfying the known phenomenological and cosmological constraints.
An explicit construction of such model is beyond the scope of this paper,
and will be the subject of upcoming work~\cite{SBM}.


\section{Conclusion}
We have examined the possibility that late decays of
massive particle after BBN can provide
a common origin for dark radiation around the epoch of recombination
and warm dark matter with free streaming which can solve the small scale structure problems.  
As a specific example, we proposed a supersymmetric axion model in which
dark radiation axions and warm dark matter axinos are produced
by the decays of saxion, and identified
a parameter space which  can successfully realize the scenario
while satisfying all the cosmological constraints.

\section*{Acknowledgments}

K.C is supported by the KRF Grants funded by the Korean Government (KRF-2008-314-C00064 and KRF-2007-341-C00010) and the KOSEF Grant funded by the Korean Government (No. 2009-0080844).
K.-Y.C and C.S.S were supported by Basic Science Research Program through the National Research Foundation of Korea (NRF) funded by the Ministry of Education, Science and Technology (No. 2011-0011083).
K.-Y.C and C.S.S acknowledges the Max Planck Society (MPG), the Korea Ministry of
Education, Science and Technology (MEST), Gyeongsangbuk-Do and Pohang
City for the support of the Independent Junior Research Group at the Asia Pacific
Center for Theoretical Physics (APCTP).




\begin{thebibliography}{99}
\bibitem{foot}  
  Using \eq{NTP} and \eq{Neffaxion}, with $\Delta \Neff\sim 1$ 
  and $\lambda_{\rm FS}\sim 1\mpc$ respectively,
  we can express the thermal axino relic density in terms of the mass of axino and  saxion, 
$\Omega^{\rm TP}_{\tilde a}h^2 \sim 0.1\, ({\rm GeV}/\ms)^5 (0.02/ (m/\ms))^4$.
For  the saxion mass smaller than GeV to satisfy the BBN constraints, 
$m/\ms \gtrsim 0.1$ is required to suppress the thermal axino relic density.
\end{thebibliography}
\end{document}